\documentclass[prx,twocolumn,showpacs,floatfix,amsmath,amssymb,longbibliographystyle,superscriptaddress]{revtex4-1}
\usepackage{amsfonts}
\usepackage{amsmath,bm}
\usepackage{mathrsfs}
\usepackage{tipa}
\usepackage{amssymb}
\usepackage{txfonts}
\usepackage{graphicx}
\usepackage{dcolumn}
\usepackage{epstopdf}
\usepackage[colorlinks,linkcolor=blue,urlcolor=blue,citecolor=blue]{hyperref}
\usepackage{multirow}
\usepackage{subfigure}
\usepackage{float}

\begin{document}

\title{Transient Higgs oscillations and high-order nonlinear light-Higgs coupling in terahertz-wave driven NbN superconductor}
\author{Z. X. Wang}
\thanks{Those authors contribute equally to this work.}
\affiliation{International Center for Quantum Materials, School of Physics, Peking University, Beijing 100871, China}

\author{J. R. Xue}
\thanks{Those authors contribute equally to this work.}
\affiliation{International Center for Quantum Materials, School of Physics, Peking University, Beijing 100871, China}

\author{H. K. Shi}
\affiliation{School of Electronic Science and Engineering, Nanjing University, Nanjing 210093, China}

\author{X. Q. Jia}
\affiliation{School of Electronic Science and Engineering, Nanjing University, Nanjing 210093, China}

\author{T. Lin}
\affiliation{International Center for Quantum Materials, School of Physics, Peking University, Beijing 100871, China}

\author{L. Y. Shi}
\affiliation{International Center for Quantum Materials, School of Physics, Peking University, Beijing 100871, China}

\author{T. Dong}
\email{taodong@pku.edu.cn}
\affiliation{International Center for Quantum Materials, School of Physics, Peking University, Beijing 100871, China}

\author{F. Wang}
\email{wangfa@pku.edu.cn}
\affiliation{International Center for Quantum Materials, School of Physics, Peking University, Beijing 100871, China}

\author{N. L. Wang}
\email{nlwang@pku.edu.cn}
\affiliation{International Center for Quantum Materials, School of Physics, Peking University, Beijing 100871, China}
\affiliation{Beijing Academy of Quantum Information Sciences, Beijing 100913, China}

\begin{abstract}

We study the nonlinear optical response in a superconducting NbN thin film with strong terahertz (THz) wave. Besides the expected third harmonic generation, we observe a new transient oscillation which softens in frequency with temperature increasing towards superconducting transition temperature $T_c$. We identify this new mode as the Higgs transient oscillation. To verify this proposal, we introduce a time-frequency resolved technique, named spectrogram for visualizing THz spectrum. The dynamic decaying behavior of the mode is observed, which is consistent with theoretical expectation about intrinsic Higgs oscillation. Moreover, a higher order nonlinear optics effect, $\emph{i.e.}$ fifth harmonic generation, has been observed for the first time, which we assign to the higher order coupling between Higgs mode and electromagnetic field.
\end{abstract}


\maketitle

With the recent development of state-of-the-art strong field terahertz (THz) spectroscopy, it becomes possible to study linear or nonlinear electromagnetic wave-matter coupling in regime that was not accessible before \cite{Orenstein2012,Giannetti2016}. Since the energy scale of THz light is in the range of meV, it can gently excite quantum materials without destroying their electronic orders. Unlike the infrared excitation at higher energy, the THz experiments provide more direct information about the low lying excitations or ground-state properties of quantum materials.

Among the most appealing applications of strong field THz electromagnetic wave is to excite and probe collective modes of matter, such as the Higgs mode in a superconductor \cite{Anderson1963,Higgs1964,Varma2002,Yuzbashyan2006,PhysRevB.84.174522,PhysRevB.87.054503,RN72,RN130,PhysRevB.92.064508,Pekker2015,Matsunaga2017,Yang2019,PhysRevB.100.054504,PhysRevB.101.184519,Chu2020,Schwarz2020,Shimano2020,Kovalev}. The Higgs mode emerges as a result of the spontaneous breaking of U(1) symmetry in the superconducting state, which is characterized by a Mexican hat-shaped free energy potential. In principle, there are two kinds of collective excitations associated with
this symmetry breaking order: a phase excitation azimuthally around the brim of the Mexican hat; and an amplitude excitation along the radial direction of the Mexican hat. For a superconductor, the phase part, being referred to as Nambu-Goldstone mode, is screened by long-range Coulomb interactions between charges and shifted up to the plasma frequency by Anderson-Higgs mechanism. The amplitude part is called Higgs mode, which is a massive mode with an energy at the superconducting energy gap 2$\Delta$. The Higgs mode in a superconductor is an analogy to the Higgs boson in particle physics \cite{Nambu1960,Varma2002,Volovik2014}.

Since the Higgs mode does not possess electric or magnetic dipole moment, it does not couple linearly to light. This limitation makes it difficult to be detected, except for some special cases, \emph{e.g}, the coexistence of the superconductivity and charge density wave order that makes the Higgs mode Raman-active and detectable by Raman scattering experiments \cite{Sooryakumar1980,Littlewood1981,Littlewood1982,Measson2014,Cea2014}. In recent years, with the development of strong field THz technology, the Higgs mode can be excited and probed by two different ways. One is to quench the superconducting system by a single-cycle THz pump pulse that brings it out of equilibrium state. Then, the order parameter starts to oscillate at the new equilibrium state with the Higgs mode frequency 2$\Delta$ that can be detected by the time domain THz probe measurement \cite{RN72,Giorgianni2019}. The other way is to drive the system periodically with a multicycle THz pulse \cite{RN130,Yang2019,Chu2020,Kovalev}. In the context of Ginzburg-Landau theory, the lowest order coupling between the Higgs mode $H$ and electromagnetic field $\textbf{A}(\omega)$ is given by $\textbf{A}^2(\omega)H$. This nonlinear coupling of light to the superconducting condensate induces oscillation of the order parameter at twice of the driving frequency $2\omega$. This nonlinear coupling can also induce higher-order third harmonic generation (THG) currents $j^{(3)}(t)$, resulting from the driven oscillations of the Higgs mode, \emph{i.e.} $j^{(3)}(3\omega)\propto \textbf{A}(\omega)H(2\omega)$. With the effective 2$\omega$ driving frequency approaching towards the energy of the Higgs mode 2$\Delta$, a resonance in the gap oscillation occurs and consequently in the THG intensity as well. This behaviour has been observed in the transmitted electric field \cite{RN130}. Although there has been argument that, in addition to the Higgs mode, the charge-density fluctuation or pair breaking also induces the THG with a similar resonant character at 2$\omega$ = 2$\Delta$ \cite{PhysRevB.93.180507,Cea2018}, subsequent polarization-resolved terahertz studies showed that the Higgs mode gives a dominant contribution to the THG far exceeding the charge-density fluctuation contribution in NbN superconductor \cite{Matsunaga2017,Tsuji2020}.

In this letter, we study the optical response in superconductor NbN under multicycle THz driving pulses. We observed a weak oscillation peak feature close to the THG mode. A red shift in energy of this new peak with increasing temperature is observed, coinciding with the softening of the intrinsic Higgs mode toward $T_c$. In light of theoretical analysis, we identify the feature as a transient Higgs mode. It means that the quench dynamics is present in the THz driving experiment. To characterize the time-evolution of the transient mode and the THG mode, we introduce a frequency-resolved optical gating (FROG) technique to analyze the data. We find that the transient Higgs mode decays faster than the THG mode, being consistent with our theoretical prediction. Furthermore, for the first time we identify a fifth harmonic generation (FHG), which we assign to the higher order coupling $\textbf{A}(\omega)^4H$ between Higgs mode $H$ and electromagnetic field $\textbf{A}(\omega)$.

\begin{figure*}[htbp]
	\centering
	\centering\includegraphics[width=14cm]{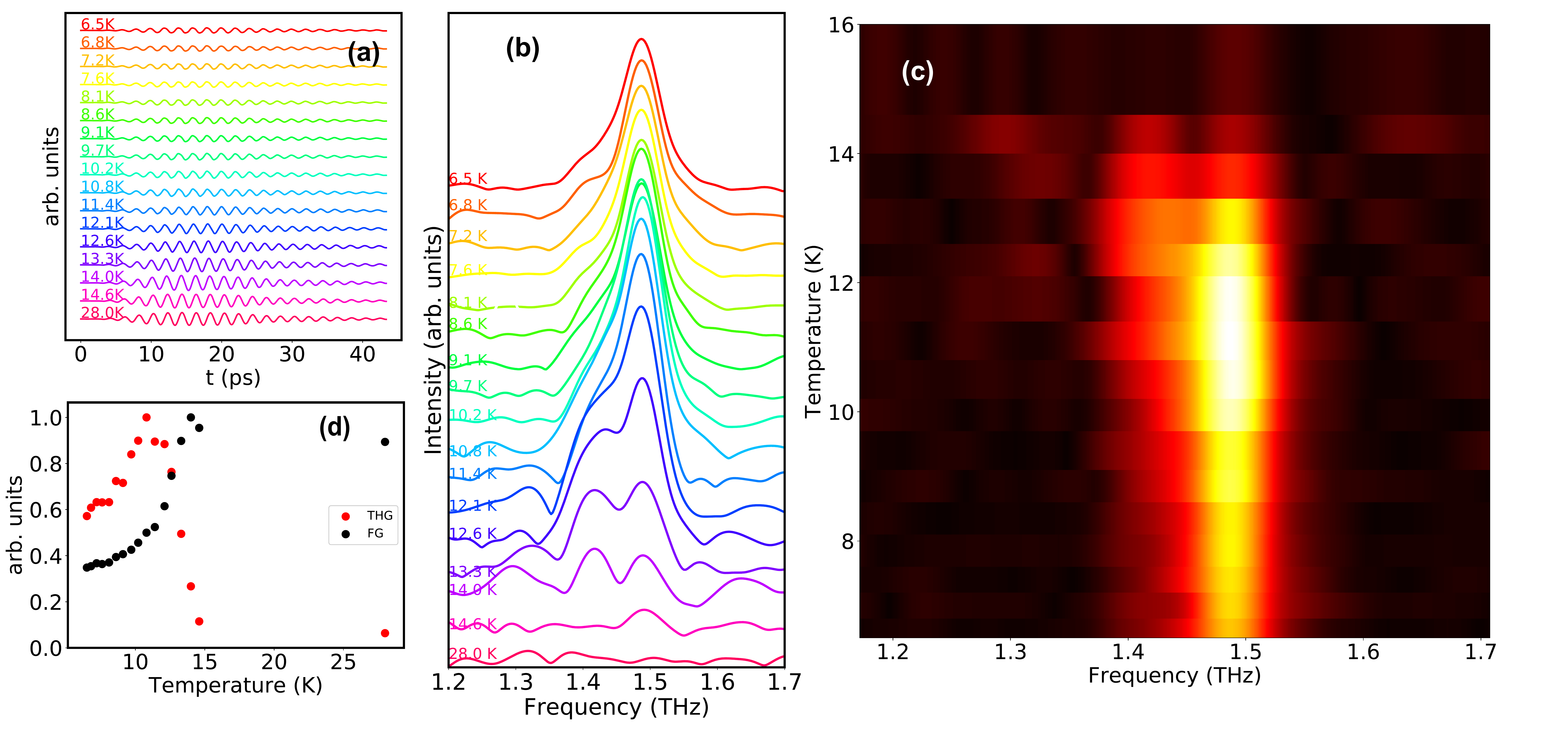}\\	
	\caption{
\textbf{Temperature dependence third harmonic generation in NbN.} (a) Strong-field THz transmission electric fields of NbN crystal ($T_{\rm{c}} = 15$ K) measured from superconducting state to normal state. (b) The spectra in frequency domain around 1.5 THz via Fourier transformation of time-domain THz curves in (a). (c) Temperature dependent intensity map in frequency domain extracted from (b). (d) Spectrum weight of THG and FG around 1.5 THz and 0.5 THz (red and black circles, respectively) versus temperature.}\label{Fig:1}
\end{figure*}

A superconducting 50-nm thickness NbN thin film grown on an MgO substrate was used for the experiment \cite{Kang2011}. It is an s-wave superconductor with transition temperature ($T_c$ = 15 K). The non-linear high harmonic generation spectra were measured in a home-built high-field terahertz transmission spectroscopy system. The output laser beam from a regenerative amplified Ti:sapphire laser system centered at 800-nm wavelength with 1-kHz repetition rate was split into two parts: one for the generation of the High-field THz by tilted-pulse-front method on a $\rm{LiNbO_3}$ crystal and the other for detecting the THz light as gating pulse for the electro-optic sampling (EOS) on ZnTe crystal. The electric field of a THz pulse that passes through the sample in cryostal is recorded as a function of the time delay between the sampling path and the THz generation path, measured by a pre-amplifier and a lock-in amplifier. Fourier transformation of the recorded time traces provides the frequency domain complex transmission spectra which contains both magnitude and phase information. More details about the light path are provided in the supplementary file \cite{Supplemental}.

\begin{figure}[htbp]
	\centering
	\includegraphics[width=8cm]{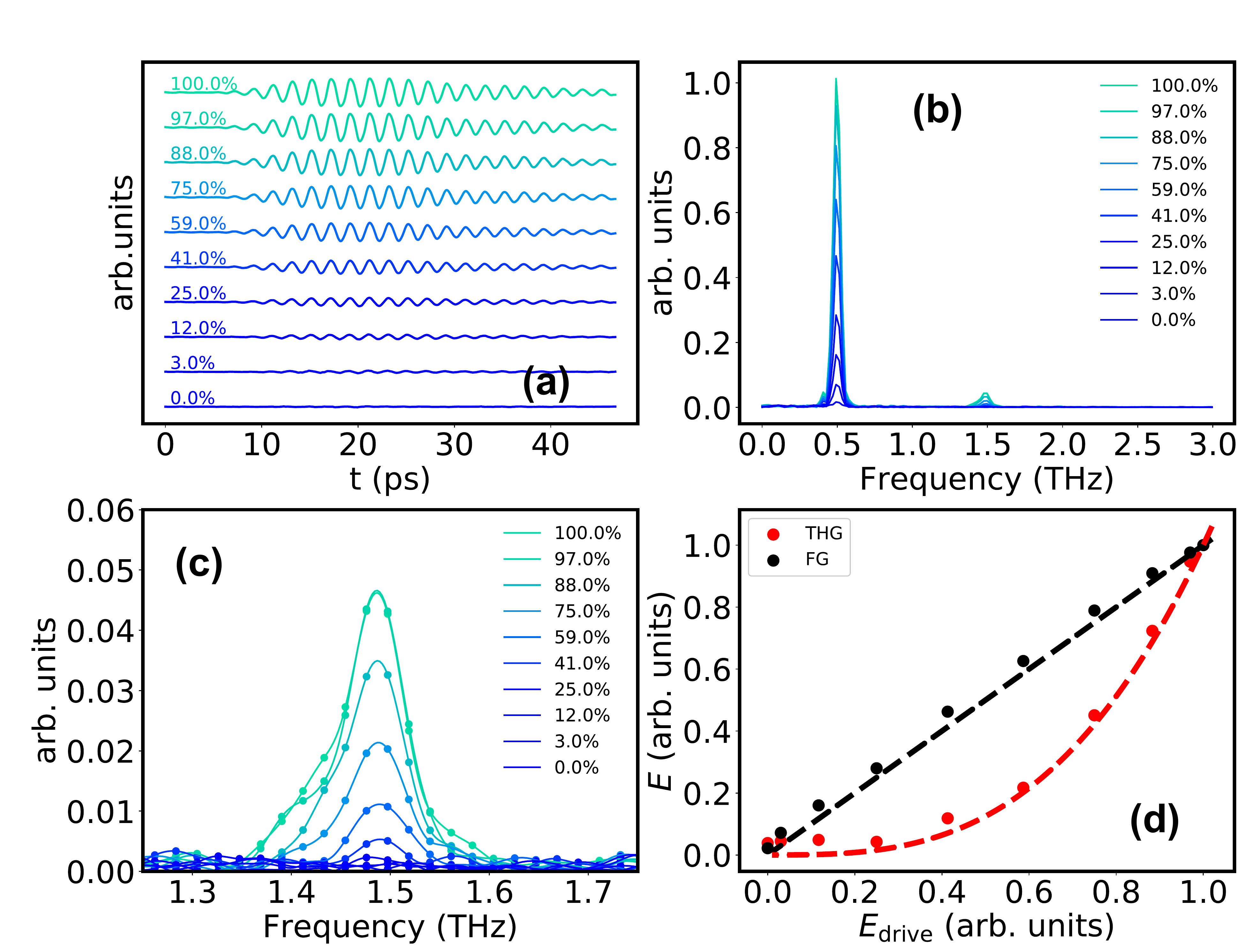}\\
	\caption{\textbf{Power dependence third harmonic generation in NbN.} (a) THz electric field transmitted through NbN at different power incoming electric field. (b) The frequency-domain spectra via Fourier transformation of (a). (c) An enlarged view of the details about THG at 1.5 THz. (d) The spectrum weight of FG and THG versus incoming electric field. The dashed black and red lines are the fitting curves by the linear and the third power function, respectively.}\label{Fig:2}
\end{figure}

To investigate the nonlinear optical response of the strong-field terahertz THG signal, we put 0.5-THz bandpass filters (BPF) before NbN film. Then, the single-cycle broadband THz light is reshaped into a multi-cycles narrow band light around 0.5 THz. We measured the time-domain THz transmission electric field of the NbN film with $T_c=15$ K by increasing temperature from superconducting state to normal state. The time traces at selected temperatures are shown in Fig.~\ref{Fig:1} (a). The Fourier transformed spectra between 1.2 - 1.7 THz in the frequency domain $|\tilde{E}(\omega)|$ are presented in Fig.~\ref{Fig:1} (b). In the low temperature superconducting state, there is a visible THG peak at 1.5 THz. As the temperature rises, the THG signal increases and reaches maximum under resonant conditions with 0.5 THz at 11.4 K. Then, it recedes and disappears above $T_c$, which confirms that the THG signal originates from superconductivity. Figure \ref{Fig:1} (c) shows the intensity map extracted from Fig.~\ref{Fig:1} (b). The THG peak at 1.5 THz does not show any shift in frequency with temperature increasing from the lowest measurement temperature at 6 K to T$_c$. Figure \ref{Fig:1} (d) shows the spectrum weight of fundamental frequency generation (FFG) and THG obtained by integration around 0.5 THz and 1.5 THz, defined as $\rm{SW(\omega_0)} = \int_{\omega_0-\Delta \omega}^{\omega_0+\Delta \omega} E(\omega)d\omega$ with $\Delta \omega=0.1$ THz. With the increase of temperature, the FFG signal monotonically increases due to the enhancement of transmittance, while the THG signal increases firstly, decreases subsequently and finally vanishes. We notice that a new discernible peak emerges as temperature rises over 10 K and a red shift of this new peak with temperature occurs, as seen in Fig.~\ref{Fig:1} (b) (c). It is attributed to the softening of the intrinsic Higgs mode and will be explained in details later.

\begin{figure*}[htbp]
	\centering
	\includegraphics[width=17cm]{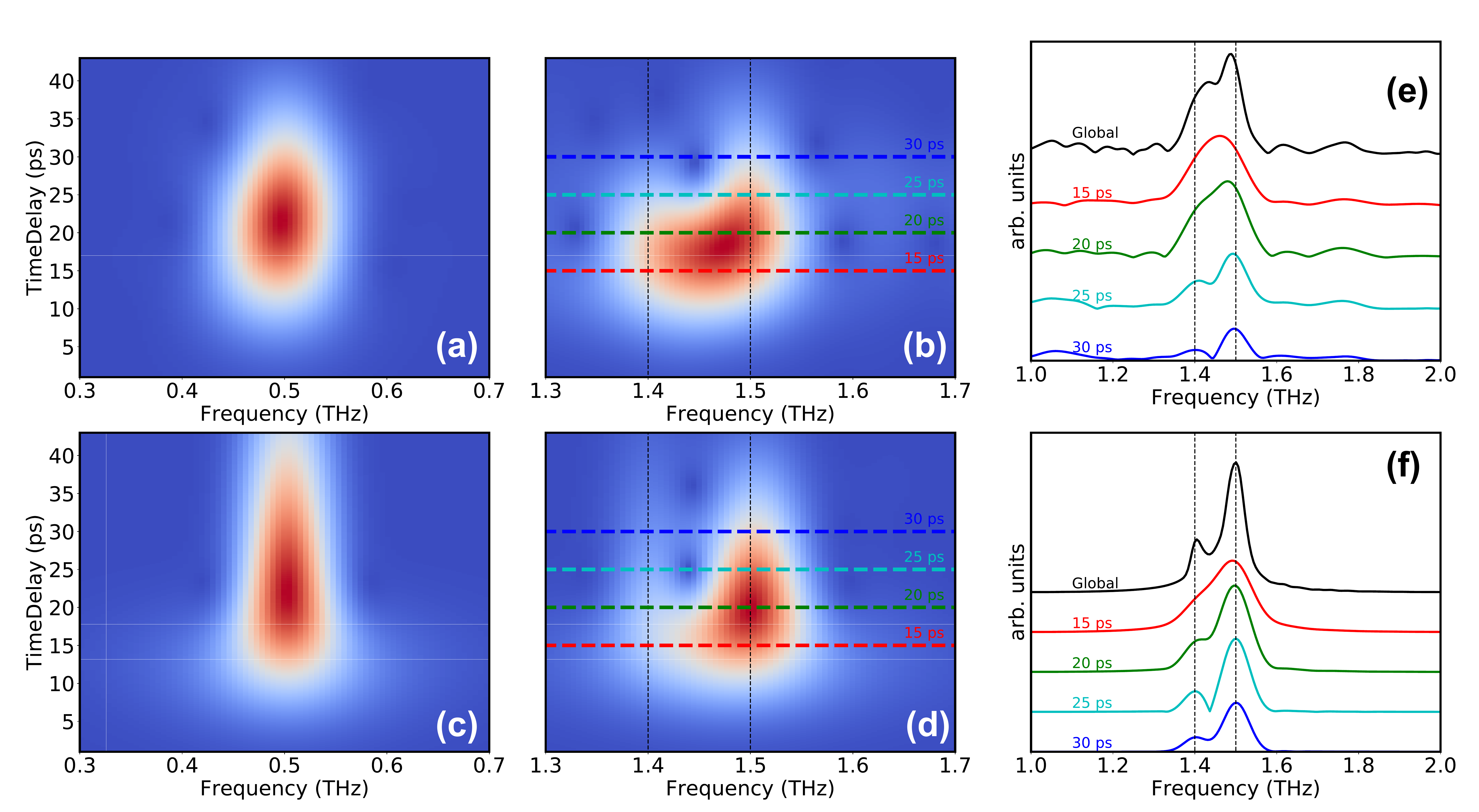}\\
	\caption{\textbf{Spectrogram traces for THG frequency-resolved optical gating (FROG).} The FROG traces are shown as contour plots, with red regions indicating higher intensity (with experimental results (a), (b) and theoretical results (c), (d)). Both traces correspond to identical Gaussian power spectra, but with different spectral phase profiles. FFG, bandwidth-linmited, as shown in (a), (c); THG, up-chirp with cubic spectral phase, as shown in (b),(d). (e), (f) The division of the Fourier transform spectrum: from red to blue lines come extracted from the dashed lines in (b) and (d).}\label{Fig:3}
\end{figure*}

As a third-order nonlinear optical process, THG is proportional to the electric field intensity to the third power. Figure \ref{Fig:2} shows the result of fluence dependent measurements at 6.5 K. We put two THz wire grid polarizers (WGP) after 0.5-THz BPF to control the intensity of incoming electric field. Due to Malus law $I \propto{|\cos^2(\theta)|}$ where $\theta$ is the phase difference between two WGPs direction, the variety of electric field can be realized by rotating the polarization direction of one THz wire grid polarizer. Figure \ref{Fig:2} (a) and (b) present time and frequency domain THz electric field $E\left( t \right)$ and $\left| {\tilde E\left( \omega  \right)} \right|$ as a function of fluence. Figure \ref{Fig:2} (c) shows the details about the THG peak at 1.5 THz. Spectrum weight of FFG and THG versus incoming THz light is shown in Fig.~\ref{Fig:2} (d), where the black and red circles are the spectrum weight extracted from integration around 0.5 THz 1.5 THz in Fig.~\ref{Fig:2} (b), respectively. The relation between THG and incoming THz light agrees well with the expected third power law: $E^{\rm{THG}}(3\omega)\propto E^3(\omega)$. The black and red dashed lines are fitting curves to the linear and third power functions, respectively.

As mentioned above, we observe a new and weak oscillation mode around 1.5 THz in Fig.~\ref{Fig:1} (b) and (c). With the temperature increasing, the frequency of the peak at 1.5 THz remains unchanged, while the peak below 1.5 THz shows red shift. We identify these two modes as the third harmonic generation and the transient Higgs mode respectively. The frequency of forced vibration is only related to the driving frequency $\omega$, thus the frequency of the THG does not depend on temperature. In contrast, the red-shifted transient Higgs mode is not only related to the unchanged driving frequency $\omega$, but also depends on the intrinsic Higgs mode energy $2\Delta(T)$ which softens with increasing temperature. When a multi-cycle strong field THz pulse arrives, the system is nonadiabatically pushed into a nonequilibrium state and will evolve towards the time-periodically forced oscillation state. At the arrival of the pump pulse, the transient Higgs mode should be predominant, and its mode frequency is temperature-dependent. As the system approaches a stable periodically driven state, the transient Higgs mode decays gradually while the THG persists and becomes predominant, whose frequency remains unchanged with a change of temperature. The transient Higgs signal disappears before the temperature reaches $T_c$ since the resonant enhancement is suppressed when $2\Delta(T)$ moves out of the narrow bandwidth of the incident pump field. The scenario proposed above is supported by our theoretical calculation based on the Anderson pseudospin precession picture, which is described in detail in Supplementary Materials \cite{Supplemental}.
We note that, in a reported work on NbN \cite{RN130}, the authors did not resolve two separate features but indicate a red shift of the THG peak when the temperature is elevated to T$_c$. The observation is somewhat different from the present work. Nevertheless, the softening of the transient Higgs mode is likely the reason for the observed red-shift of THG, which might be caused by the limited frequency resolution.

\begin{figure}[htbp]
	\centering
	\includegraphics[width=8cm]{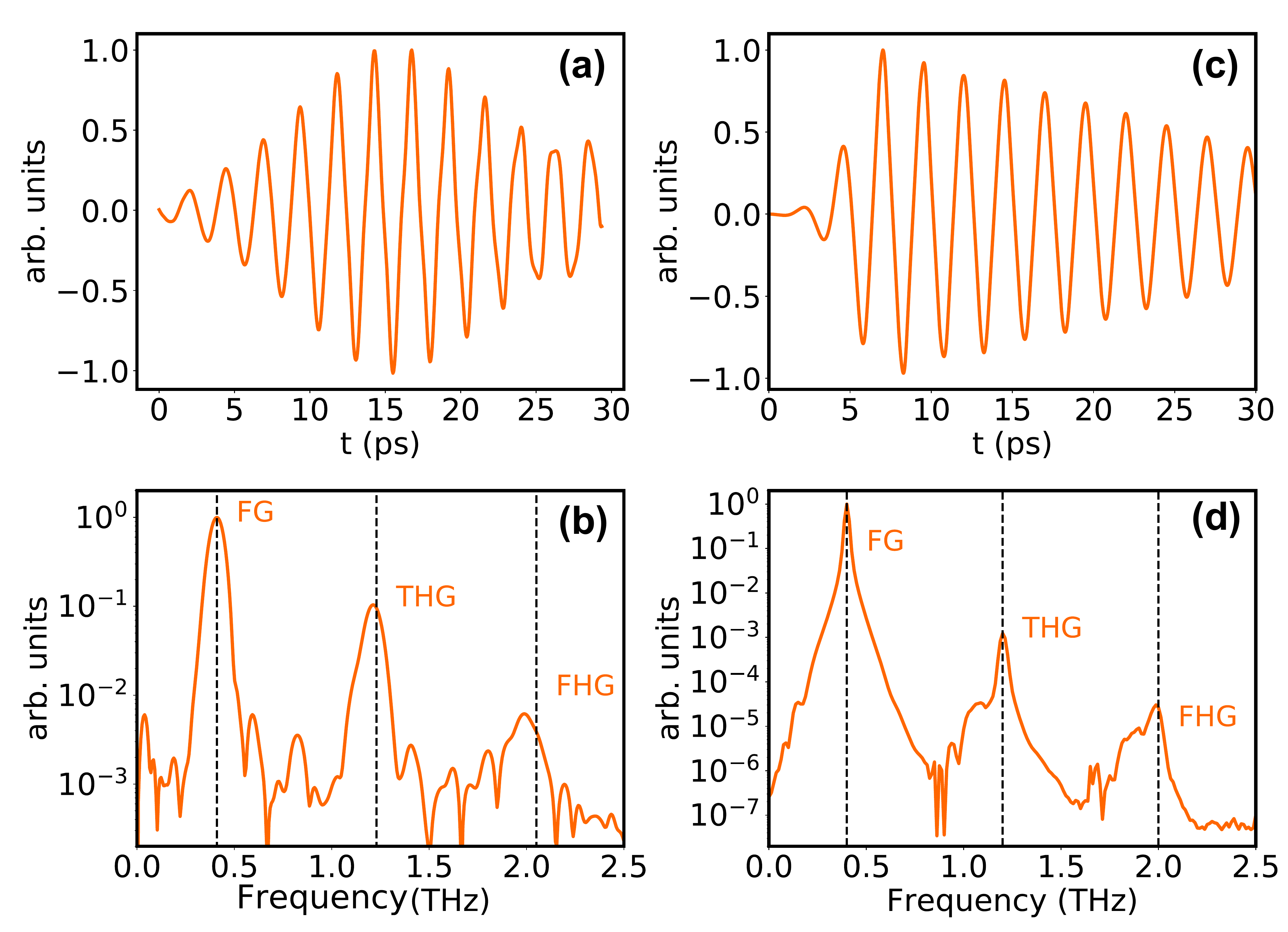}\\
	\caption{\textbf{The fifth harmonic generation in NbN.} (a) (b) Strong-field and 0.4-THz field transmitted through NbN at 6.8 K. A 0.4 THz bandpass filter is placed before sample. (c) (d)
	Similar results from theoretical calculations.}\label{Fig:4}
\end{figure}

To clarify the above proposal, we introduce the time-frequency distribution spectrogram to separate the transient Higgs mode and THG mode (more details in supplementary \cite{Supplemental}). Spectrogram is an analysis tool being used to characterize signals whose spectral content is varying in time \cite{Weiner_2009}. The techniques are called frequency-resolved optical gating, also known by the acronym FROG. It is widely used in femtosecond laser pulse measurement, and FROG trace visually displays the frequency versus time delay. Here, we use it to analyze the time domain terahertz spectrum. The time-frequency resolved result is extracted from time-domain spectrum with different time delay gating windows by
\begin{equation}
S(\omega,\tau)\equiv\left|\int ^{\infty}_{-\infty}E(t)g(t-\tau)\textrm{exp}(-\textbf{i}\omega t)\textrm{d}t\right|^2
\end{equation}
where the gating function is set as Gaussian function $g(t-\tau)=\textrm{e}^{-[(t-\tau)/T]^2}$. \emph{T} is the time window used in the integration. We choose \emph{T}=6 ps in the integration. We checked that the effect is small if a different value is used for the time window. Figure \ref{Fig:3} shows the time-frequency spectrogram derived from time domain THz spectrum at 12.6 K. The experimental and theoretical results are as shown in Fig.~\ref{Fig:3} (a), (b) and (c), (d), respectively (See supplementary for details \cite{Supplemental}). We notice that the main peak at 0.5 THz is free of phase or frequency modulation as shown in Fig.~\ref{Fig:3} (a) and (c). If the signal around 1.5 THz was entirely THG and no other components, it should also be free of phase or frequency modulation, the same as the main peak at 0.5 THz. However, a positive chirp, \emph{i.e.} longer times correspond to higher frequencies in THG is observed, as shown in Fig.~\ref{Fig:3} (b) and (d). It demonstrates that there is another non-linear optical response besides THG, that is the transient Higgs mode. The theoretical and experimental results are very similar. The frequency of the transient Higgs mode and THG is indicated in the Fig.~\ref{Fig:3} (b) and (d) by dotted black lines (the temperature mismatch in experiment and theoretical calculation may be due to the too simple model we adopt to simulate the time evolution). The spectrum can be divided into two peaks by different delay time $\tau$ of $g(t-\tau)$, shown in Fig.~\ref{Fig:3} (b) and (d) by the dashed lines from blue to red. The global and frequency domain spectrua with different gating windows are shown in Fig.~\ref{Fig:3} (e) experimentally, (f) theoretically. The transient Higgs mode decays faster than THG. Since the transient Higgs mode is observed in the driving experiment, the results may provide additional support that the THG signal is contributed dominantly from the Higgs excitation rather than the charge density fluctuation.

Finally, we would like to present another striking and interesting observation in our THz pulse driving measurement. We observed a weak high-order (fifth order) nonlinear optics effect when we placed 0.4-THz bandpass filters before NbN sample. Here, a multi-cycle narrow band 0.4 THz pulse is generated and employed as the fundamental frequency light. Figure~\ref{Fig:4} (a) shows the measured time-domain THz transmission electric field of the NbN film at 6.8 K in superconducting state. The Fourier transformed spectrum is shown in Fig.~\ref{Fig:4} (b). We observed a weak but well-recognizable FHG peak around 2.0 THz in addition to the stronger THG signal at 1.2 THz. As the temperature rises, the THG and FHG signals further weaken and totally vanish when the sample goes into normal state. Such higher order harmonic generation was not observed/reported before. We also observed a similar
FHG with the driving fundamental frequency of 0.3 THz, but much weaker strength (Result is shown in the Supplementary Materials \cite{Supplemental}). However, we can not resolve such FHG signal when the driving fundamental frequency is 0.5 THz at 6.8 K. Since the frequencies of FHG for 0.3, 0.4 and 0.5 THz incoming driving wave are 1.5, 2.0 and 2.5 THz, respectively, the relatively strong FHG signal is likely linked to the fact that 2.0 THz is more close to the resonant energy of 2$\Delta$. We assign FHG to the higher order coupling $\textbf{A}^4(\omega)H$ between Higgs mode $H$ and electromagnetic field $\textbf{A}(\omega)$.
This coupling will induce driven oscillations of the Higgs mode at frequency $4\omega$ whose amplitude will be largest when $4\omega$ matches the intrinsic Higgs mode energy $2\Delta(T)$, and it will further produce a FHG component of current $j^{(5)}(5\omega)\propto \textbf{A}(\omega) H(4\omega)$.
Our theoretical modeling can indeed reproduce the observed FHG result. Detailed information is presented in the Supplementary Materials \cite{Supplemental}.

To summarize, we observe a new transient Higgs mode around the THG mode and a red-shift behavior for the new transient mode as temperature increases. We propose a scenario that a quench dynamics is still present in the driving experiment to explain this new mode. We confirm this assumption by time-frequency distribution spectrogram. The temperature-dependence signal decays fast, while the strict THG signal at $3\omega$ oscillates at much longer time delay. We also observe a weak but clearly discernable fifth-harmonic generation in thin-film NbN sample, which we assign to the higher order coupling between Higgs mode and electromagnetic field.

\begin{center}
\small{\textbf{ACKNOWLEDGMENTS}}
\end{center}
This work was supported by National Natural Science Foundation of China (No. 11888101), the National Key Research and Development Program of China (No. 2017YFA0302904).

\nocite{*}
\bibliography{NbN}

\end{document}